\documentclass[12pt]{iopart}
\pdfoutput=1
\expandafter\let\csname equation*\endcsname\relax
\expandafter\let\csname endequation*\endcsname\relax
\usepackage{color}
\usepackage[breakwords]{truncate}
\usepackage[font={small}]{caption}
\usepackage{physics}
\usepackage{graphicx}
\usepackage{cite} 
\usepackage{subcaption}
\usepackage{etoolbox}
\usepackage[section]{placeins}
\usepackage[sectionbib]{chapterbib}
\usepackage{amsfonts}
\usepackage[citecolor=magenta,colorlinks=true]{hyperref}
\makeatletter
\newrobustcmd{\fixappendix}{%
  \patchcmd{\l@section}{1.5em}{7em}{}{}%
  \patchcmd{\l@subsection}{2.3em}{7em}{}{}%
}
\makeatother
\appto\appendix{
\addtocontents{toc}{\fixappendix}
\addtocontents{toc}{\protect\setcounter{tocdepth}{1}}}

\begin{document}
\newcommand{\ben}[1]{\textcolor{blue}{\textbf{#1}}}
\title[First passage statistics of Poisson random walks]{First passage statistics of Poisson random walks on lattices}
\author{Stephy Jose}
\address{ Tata Institute of Fundamental Research, Hyderabad, India}
\eads{\mailto{stephyjose@tifrh.res.in}}
\begin{abstract}
The first passage statistics of a continuous time random walker with Poisson distributed jumps on one and two dimensional infinite lattices is investigated. An exact expression for the probability of first return to the origin in one dimension is derived for a symmetric random walker as well as a biased random walker. The Laplace transform of the occupation probability of a site for a symmetric random walker on a two dimensional lattice is identified with the lattice Green's function for a square lattice. This allows computation of the exact first passage distribution to any arbitrary site on the square lattice in Laplace space. All analytical results are compared with kinetic Monte Carlo simulations of a lattice walker in one and two dimensions. 
\end{abstract}

\newpage
{\pagestyle{plain}
 \tableofcontents
\cleardoublepage}
\section{Introduction}
\label{intro}
The classic problem of first passage statistics has long been of interest due to its applications in diverse fields such as biology~\cite{chou2014first,kenwright2012first,condamin2008probing}, reactions~\cite{szabo1980first,park2003reaction}, and trade triggers~\cite{liu2017anchoring,zhang2009first,chicheportiche2014some} amongst others. George Polya's famous recurrence theorem~\cite{polya1921aufgabe} states that a symmetric discrete time random walk is recurrent in one and two dimensions, but transient in three dimensions. However, a biased random walk is transient in all dimensions where the probability of ever returning to the origin is always less than $1$. Polya studied the symmetric random walk problem in discrete space and discrete time. In this paper, we analyse the same problem in continuous time. Continuous Time Random Walks (CTRWs)~\cite{montroll1965random,montroll1979enriched} find a variety of applications in diverse fields such as financial economics, fractional calculus (anomalous diffusion) as well as queueing theory~\cite{kutner2017continuous,mainardi2020advent}.

There have been numerous studies on first passage problems in stochastic processes~\cite{siegert1951first,burkhardt2014first,benichou2014first,khantha1983first,balakrishnan1983first}. Discussions on the asymptotics of the first return probability for discrete time random walks on  lattices in different dimensions can be found in~\cite{redner2001guide}. The first passage properties of a biased random walker in one dimension in the continuous time domain had been extensively analysed in~\cite{feller2008introduction,balakrishnan1983some}. These studies are for the first passage to any site other than the origin. The exact expression for the probability of first return to the origin of a biased continuous time random walker in one dimension is still not derived to the best of the author's knowledge.

Although the exact form for the occupation probability of a lattice site for a continuous time random walker with Poisson distributed jumps on a $d$ dimensional lattice is exactly known~\cite{balakrishnan1983some,balakrishnan1988first,balakrishnan1981two,prasad1984biased}, the exact expression for the Laplace transform of the occupation probability is not known even in two dimensions. Since there are fundamental relations connecting the occupation probability and first passage probability distributions in the Laplace domain, a closed form expression for the Laplace transform of the occupation probability gives direct insight into the first passage probability distribution.

In this paper, we analyse the first passage statistics of a random walker on one and two dimensional lattices with nearest neighbour jumps.
Time is treated as a continuous variable. The instants at which the jumps occur are Poisson distributed and the process considered is Markovian. First, the exact closed form expression for the first return probability to the starting site for a biased random walker in one dimension is derived in terms of modified Bessel functions and Struve functions. Next, an exact expression for the characteristic function of the first passage probability of a symmetric random walker to an arbitrary site in two dimensions is derived in terms of generalised hypergeometric function. In general, characteristic functions are useful as the term by term inversion of the series expansion of the characteristic function in the transformed variable in the required limit gives the corresponding limiting forms in the real domain. For example, one could extract the asymptotics of the time dependent quantities by knowing the corresponding Laplace transforms. As the Laplace transform is in fact the moment generating function, we can compute all the moments of the corresponding distribution. Thus, one  can calculate the mean first passage time as well as the recurrence properties of the random walker. The exact characteristic function for the first passage probability density of a symmetric random walker on a two dimensional square lattice is derived by mapping to the square lattice Green's function. In addition, the already known results for the probability distributions such as the occupation probability of a general site in one and two dimensions and the probability of first passage to any site other than the origin in one dimension are also derived in this paper for completeness.

\section{ Biased random walk in one dimension}
\label{sec:2}
We first consider the motion of a biased random walker on an infinite one dimensional lattice where the lattice points are labelled by the variable $x$; $x$ can take any integer values. The simple case where the walker is allowed to hop only to the nearest neighbour sites is studied. The translational rates for a positively directed walker along the positive and negative $x$ directions are  $\left(\frac{1}{2}+\epsilon \right)$ and $\left(\frac{1}{2}-\epsilon \right)$ respectively. Similarly, the translational rates for a negatively directed walker along the positive and negative $x$ directions are  $\left(\frac{1}{2}-\epsilon \right)$ and $\left(\frac{1}{2}+\epsilon \right)$ respectively. The value of $\epsilon$ is bounded between $0$ and $\frac{1}{2}$. Let us define $P(x,t)$ as the probability  for a biased random walker to be at site~$x$ at time~$t$ subjected to the initial condition that $P(x,0)=\delta_{x,0}$. One can write the equation for the evolution of $P(x,t)$ as
\begin{small}
\begin{equation}
\label{eq:a1}
 \frac{\partial P(x,t)}{\partial t}=\left(\frac{1}{2}+\epsilon \right)P(x-1,t)+\left(\frac{1}{2}-\epsilon \right)P(x+1,t)-P(x,t),
\end{equation}
\end{small}
where $\left(\frac{1}{2}+\epsilon \right)$ and $\left(\frac{1}{2}-\epsilon \right)$ are the probabilities to hop to right and left respectively. This equation holds for a walker biased along the positive $x$ direction. One can write a similar equation for a walker biased along the negative $x$ direction.
\subsection{Occupation probability}
\label{subsection:2:1}
The Laplace transform of $P(x,t)$ is defined as
\begin{equation}
 \tilde P(x,s)=\int_{0}^{\infty}e^{-st}P(x,t)dt.
\end{equation}
Laplace transforming equation (\ref{eq:a1}) gives
\begin{small}
\begin{eqnarray}
\label{eq:a2}
 P(x,0)&=&(s+1)\tilde P(x,s)-\left(\frac{1}{2}+\epsilon \right)\tilde P(x-1,s)-\left(\frac{1}{2}-\epsilon \right)\tilde P(x+1,s).
\end{eqnarray}
\end{small}
Solving (\ref{eq:a2}) for the initial condition $P(x,0)=\delta_{x,0}$ yields
\begin{equation}
\label{eq:a3}
 \tilde P(x,s)=\begin{cases}
\frac{1}{\sqrt{s(s+2)+4 \epsilon^2}}{\left(\frac{(s+1)-\sqrt{s(s+2)+4 \epsilon^2}}{(1-2 \epsilon)}\right)}^x,& x\geq 0, \\
\frac{1}{\sqrt{s(s+2)+4 \epsilon^2}}{\left(\frac{(s+1)+\sqrt{s(s+2)+4 \epsilon^2}}{(1-2 \epsilon)}\right)}^x, & x\leq 0 .
\end{cases}
\end{equation} 
For $x=0$ we obtain,
\begin{equation}
\label{0}
\tilde P(0,s)=\frac{1}{\sqrt{s(s+2)+4 \epsilon^2}}.
\end{equation}
These Laplace transforms can be easily inverted and one gets the known result~\cite{feller2008introduction,balakrishnan1983some}:
\begin{equation}
\label{eq:a3a}
P(x,t)=e^{-t}\sqrt{{\left(\frac{(1+2 \epsilon)}{(1-2 \epsilon)}\right)}^{x}}I_{\left |x  \right |}( \sqrt{1-4 \epsilon ^2}t),~\forall~x \in \mathbb{Z},
\end{equation}
where~$I_n(z)$ is the modified Bessel function of order $n$ which is a solution to the homogeneous Bessel differential equation $z^2\frac{d^2y}{dz^2}+z\frac{dy}{dz}-(z^2+n^2)y=0$. Thus we get the occupation probability of the origin as
\begin{equation}
\label{eq:a3ab}
P(0,t)=e^{-t}I_0( \sqrt{1-4 \epsilon ^2}t)=e^{-t}I_0( 2\sqrt{\alpha}t),
\end{equation}
where $\alpha=\frac{1}{4}-\epsilon^2$.
The limits of the distribution are
\begin{small}
\begin{equation} 
\label{symas2B}
\lim_{t \rightarrow 0} P(0,t)=1-t+\left(\frac{1}{2}+\alpha \right) t^2-\left(\frac{1}{6}+\alpha \right) t^3+...
\end{equation}
\end{small}
and
\begin{small}
\begin{eqnarray} 
\label{symas2C}
\lim_{t \rightarrow \infty} P(0,t)&=& \frac{e^{(-1+2\sqrt{\alpha} ) t}}{2\sqrt{\pi } \alpha^{\frac{1}{4} }} \frac{1}{\sqrt{t}}+\frac{e^{(-1+2\sqrt{\alpha} ) t}}{32\sqrt{\pi } \alpha^{\frac{3}{4} }} \frac{1}{t^{\frac{3}{2}}}+\frac{9e^{(-1+2\sqrt{\alpha} ) t}}{1024\sqrt{\pi } \alpha^{\frac{5}{4} }} \frac{1}{t^{\frac{5}{2}}}+...
\end{eqnarray}
\end{small}

\begin{figure}[t!]
    \centering
    \hspace{-1.9 cm}
    \begin{subfigure}[t]{0.44\textwidth}
        \centering
        \includegraphics[width=\textwidth]{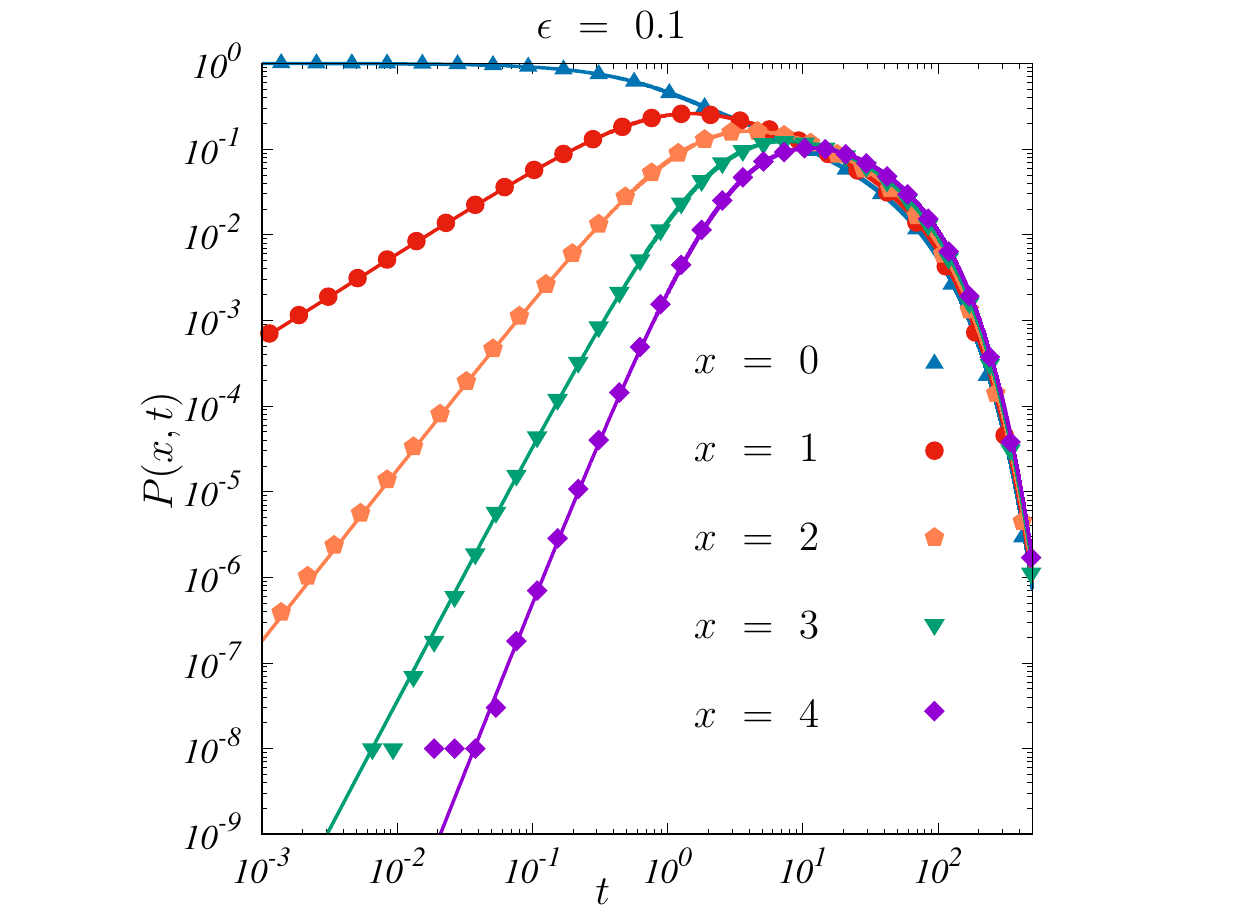}
        \caption{ }
    \end{subfigure}\hspace{-1.9 cm}
    \begin{subfigure}[t]{0.44\textwidth}
        \centering
        \includegraphics[width=\textwidth]{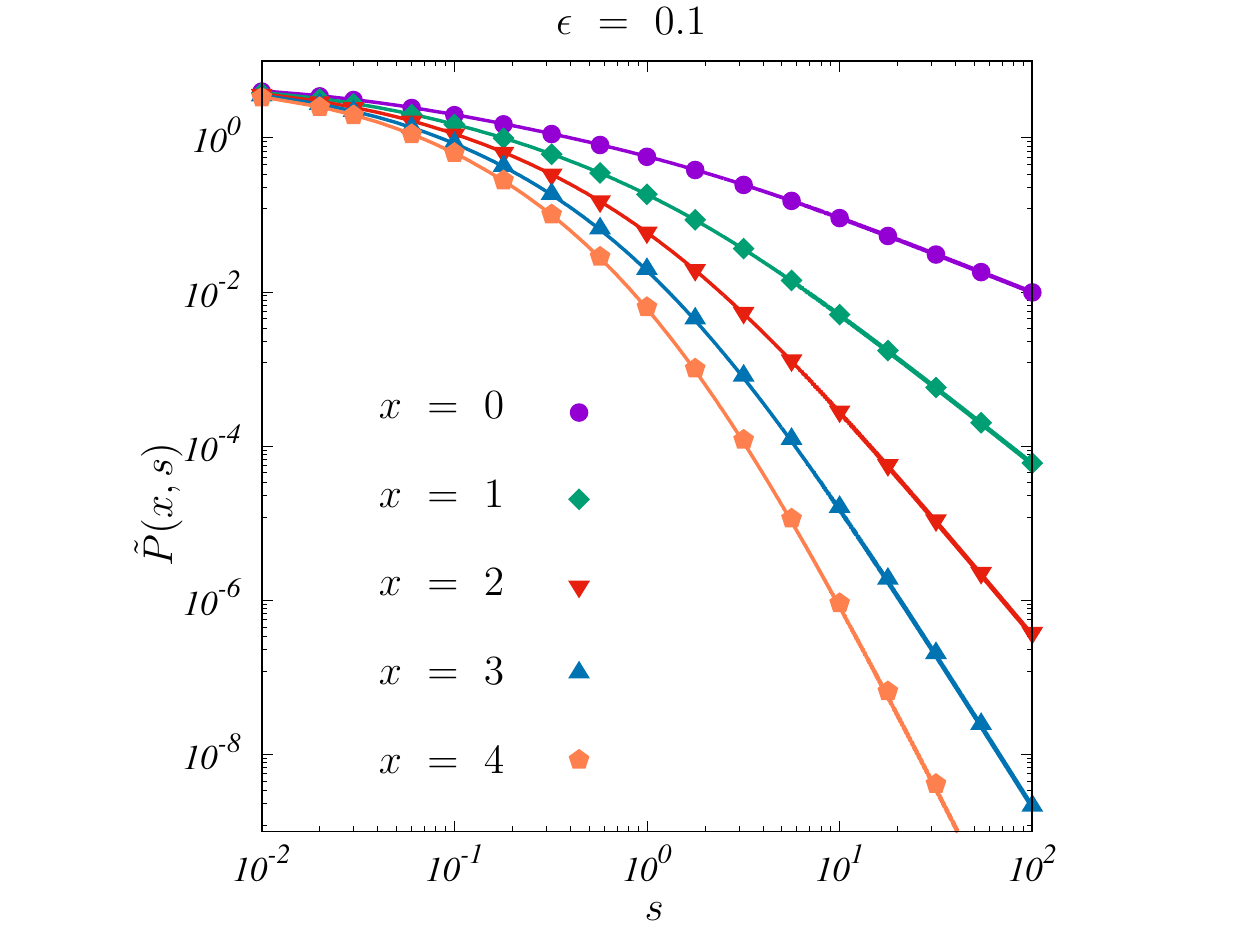}
        \caption{ }
    \end{subfigure}\hspace{-1.9 cm}
    \begin{subfigure}[t]{0.44\textwidth}
        \centering
        \includegraphics[width=\textwidth]{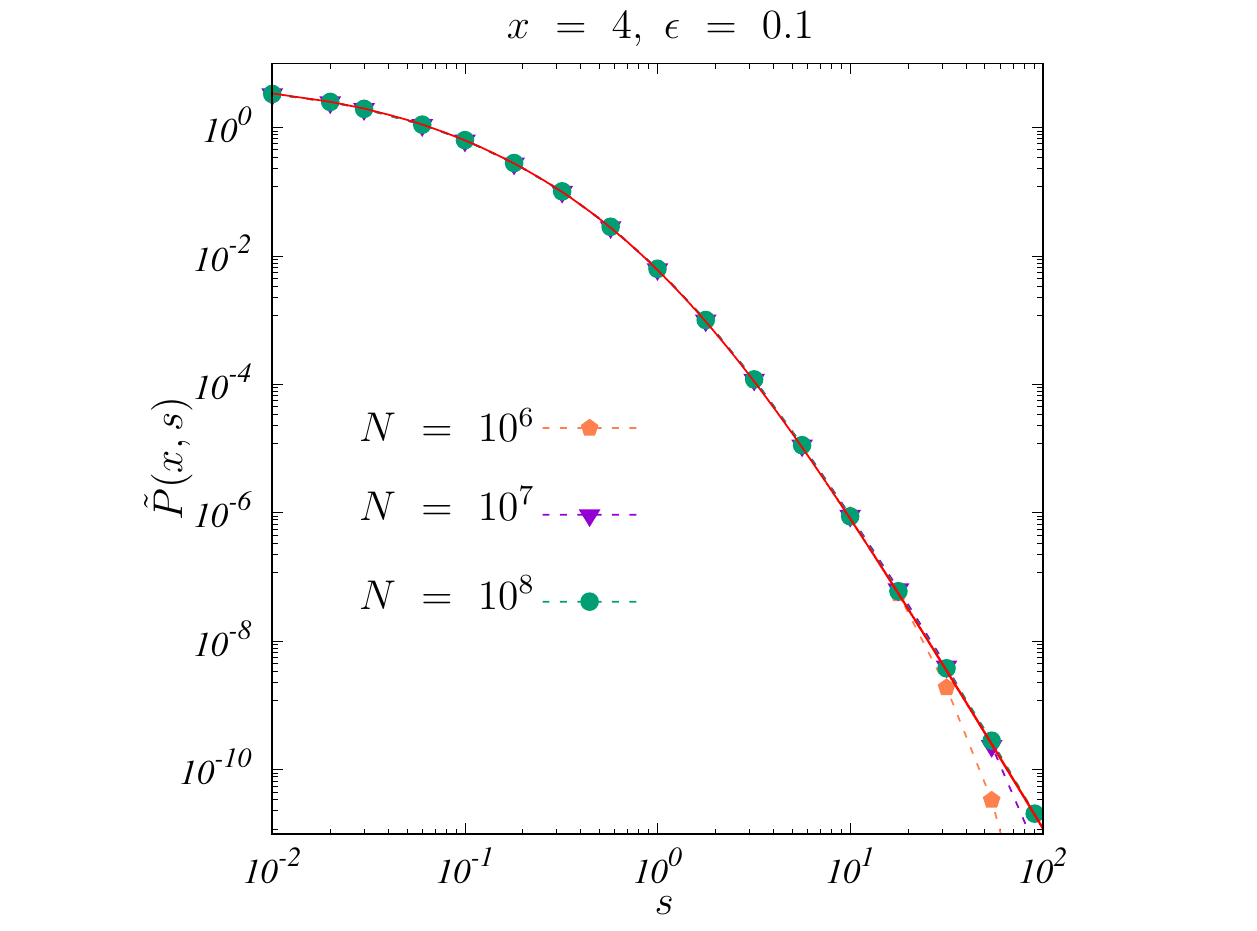}
        \caption{ }
    \end{subfigure}
    \hspace{-2 cm}
    \caption{(a)~Occupation probability of a biased random walker $P(x,t)$, plotted as a function of time for different $x$. The solid lines correspond to the theoretical result in (\ref{eq:a3a}) whereas the points are from simulations. The fixed bias used is $\epsilon=0.1$. To sample the small probabilities at short times, averaging has to be done over many realisations. Here, the simulation data is averaged over $10^8$ realisations.~(b)~The Laplace transform of the occupation probability of a biased random walker $\tilde P(x,s)$,  plotted as a function of $s$ for different $x$. The solid lines correspond to the theoretical result in (\ref{eq:a3}) and the points are from simulations.~(c) Numerical simulation data (points) for $\tilde P(x,s)$ converge with the theoretical result (red solid line) more accurately as the number of realisations $N$, increases. The error in $P(x,t)$ due to finite sampling at short times appears as a correction to $\tilde P(x,s)$ in the large $s$ limit. Here, $x$ is fixed to be $4$.}
   \label{fig:a1px}
\end{figure}
 For a symmetric random walk ($\epsilon=0$, $\alpha=\frac{1}{4}$), (\ref{eq:a3ab}) reduces to\\
\begin{equation}
\label{eq:a3b}
{P(0,t)}_{srw}=e^{-t}I_0(t).
\end{equation}
The limits are given as
\begin{equation} 
\label{symas2BS}
\lim_{t \rightarrow 0} {P(0,t)}_{srw}\approx 1-t+\frac{3}{4}  t^2-\frac{5}{12} t^3+...
\end{equation}
and
\begin{equation} 
\label{symas2CS}
\lim_{t \rightarrow \infty} {P(0,t)}_{srw}\approx \frac{1}{\sqrt{2\pi }} \frac{1}{\sqrt{t}}+\frac{1}{8\sqrt{2\pi }} \frac{1}{t^{\frac{3}{2}}}+\frac{9}{128\sqrt{2\pi }} \frac{1}{t^{\frac{5}{2}}}+...
\end{equation}

The closed form expressions for the site occupation probability~$P(x,t)$ and the corresponding Laplace transform~$\tilde P(x,s)$, in~(\ref{eq:a3a}) and (\ref{eq:a3}) are compared with the kinetic Monte Carlo (kMC) simulation results in figure~\ref{fig:a1px}. The kinetic Monte Carlo algorithm~\cite{prados1997dynamical,bortz1975new,voter2007introduction} allows efficient simulations of processes with arbitrary holding times by creating event-to-event transitions directly. Logarithmic binning is done to sample the small probabilities at short times. To get better statistics at short times, sampling has to be done over many realisations. Numerically,~$\tilde P(x,s)$ is obtained by a numerical Laplace transform of the data for~$P(x,t)$. The error in $P(x,t)$ due to finite sampling at short times appears as a correction to $\tilde P(x,s)$ in the large $s$ limit as seen from figure~\ref{fig:a1px}. This error decreases as we increase the number of realisations being averaged.
\subsection{First passage probability}
\label{subsection:2:2}
Let us define the first passage probability density $F(x,t)$,  as the probability that a random walker starting from the origin ($x=0$) at time $t=0$ arrives at $x$ for the first time at $t$. The  first passage probability density $F(x,t)$ is related to the site occupation probability $P(x,t)$ through
\begin{equation}
\label{eq:a9a}
P(x,t)=\int_{0}^{t}F(x,t')P(0,t-t')dt'+\delta_{x,0}e^{-t}.
\end{equation}
The first term in the equation accounts for the walks that first reach $x$ at time $t' \leq t$ and then return to $x$ in the remaining interval of time $t-t'$ \cite{haus1987diffusion}. The second term accounts for the initial condition as well as the persistence of the walker in the initial position. 
Taking a Laplace transform of the above equation gives
\begin{equation}
\label{fxs}
\tilde F(x,s)=\frac{\tilde P(x,s)}{\tilde P(0,s)},~\forall~x \neq 0,
\end{equation}
and
\begin{equation}
\label{rec}
\tilde F(x=0,s)=\tilde F(0,s)=1-\frac{1}{\tilde P(0,s)(s+1)}.
\end{equation}
The  probability of return to the origin upto time $t$, $R(0,t)$, is defined as
\begin{equation}
R(0,t)=\int_{0}^{t}F(0,{t}^{'}){dt}^{'},
\end{equation}
and the probability $R$, of ever returning to the origin  is defined as
\begin{equation}
R=\tilde F (0,0)=\int_{0}^{\infty}F(0,t)dt=R(0,\infty).
\end{equation}
By definition,
\begin{equation} 
F(0,t=0)=R(0,t=0)=0.
\end{equation}

Using (\ref{eq:a3}) in (\ref{fxs}) gives
\begin{equation}
\label{eq:a4}
 \tilde F(x,s)=\begin{cases}
{\left(\frac{(1+2 \epsilon)}{(s+1)+\sqrt{s(s+2)+4 \epsilon^2}}\right)}^x,& x> 0, \\
{\left(\frac{(1-2 \epsilon)}{(s+1)+\sqrt{s(s+2)+4 \epsilon^2}}\right)}^{|x|}, & x< 0,
\end{cases}
\end{equation} 
which on inverting gives the well known result~\cite{balakrishnan1983some}:
\begin{equation}
\label{eq:a5}
F(x,t)=e^{-t}\sqrt{{\left(\frac{(1+2 \epsilon)}{(1-2 \epsilon)}\right)}^{x}}\frac{{|x|}}{t}I_{|x|}(\sqrt{1-4 \epsilon^2}t),~x \neq 0.
\end{equation}
For the symmetric random walk in one dimension, this equation reduces to
\begin{equation}
\label{eq:a5a}
{F(x,t)}_{srw}=e^{-t}\frac{{|x|}}{t}I_{|x|}(t),~x \neq 0.
\end{equation}
 Let us now try to derive an exact expression for the first return probability to the starting site ($x=0$).
 Using (\ref{0}) in (\ref{rec}) gives
\begin{equation}
\label{eq:a8}
\tilde F (0,s)=1-\sqrt{1-\frac{(1-4 \epsilon^2)}{{(s+1)}^2}}.
\end{equation}
The above equation can be alternatively derived as follows: To study the return probability to the origin, one has to take account of the fact that the particle has waited at the origin for some time $\tau_r$ (which is the residual time of the walker) before making the first jump. One can then write a recursion relation for the first return probability to the origin in terms of the first passage probabilities to $x=+1$ and $x=-1$.
\begin{equation}
\label{eq:a6}
F(0,t|\tau_r)= \left(\frac{1}{2}-\epsilon \right)F(+1,t-\tau_r)+\left(\frac{1}{2}+\epsilon \right)F(-1,t-\tau_r),
\end{equation}
where $F(0,t|\tau_r)$ is the conditional first return probability density. The condition is that the first return to the origin at time $t$ happens only after waiting for time $\tau_r$ before the first jump. If the walker chooses to hop to right (happens with probability $\frac{1}{2}+\epsilon $ for a positively biased walker) after waiting for time $\tau_r$ at the origin, one can shift the origin to $x=+1$ and say that the first return probability to the origin at time $t$ is same as the first passage probability to $x=-1$ in the remaining time $t-\tau_r$ from the new shifted origin. If the walker makes a left hop (happens with probability $\frac{1}{2}-\epsilon $), then the new shifted origin is $x=-1$ and the first return probability to the origin at time $t$ is same as the first passage probability to $x=+1$ in the remaining time $t-\tau_r$ from this shifted origin. \\
From Bayes' theorem, the first return probability density at time $t$ is 
\begin{equation}
\label{eq:a7}
F(0,t)=\int_{0}^{t}F(0,t|\tau_r)P(\tau_r)d \tau_r,
\end{equation}
 where $P(\tau_r)=e^{-\tau_r}$ is the Poisson waiting time distribution. Laplace transforming (\ref{eq:a7}) and using (\ref{eq:a4}) for $F(+1,s)$ and $F(-1,s)$ gives (\ref{eq:a8}).

The probability $R$, of ever returning to the origin for a biased random walker can be found as
\begin{equation}
R=\tilde F (0,0)=\int_{0}^{\infty}F(0,t)dt=1-2 \epsilon <1.
\end{equation}
\begin{figure}[t!]
    \centering
    \hspace{-1.9 cm}
        \begin{subfigure}[t]{0.44\textwidth}
        \centering
        \includegraphics[width=\textwidth]{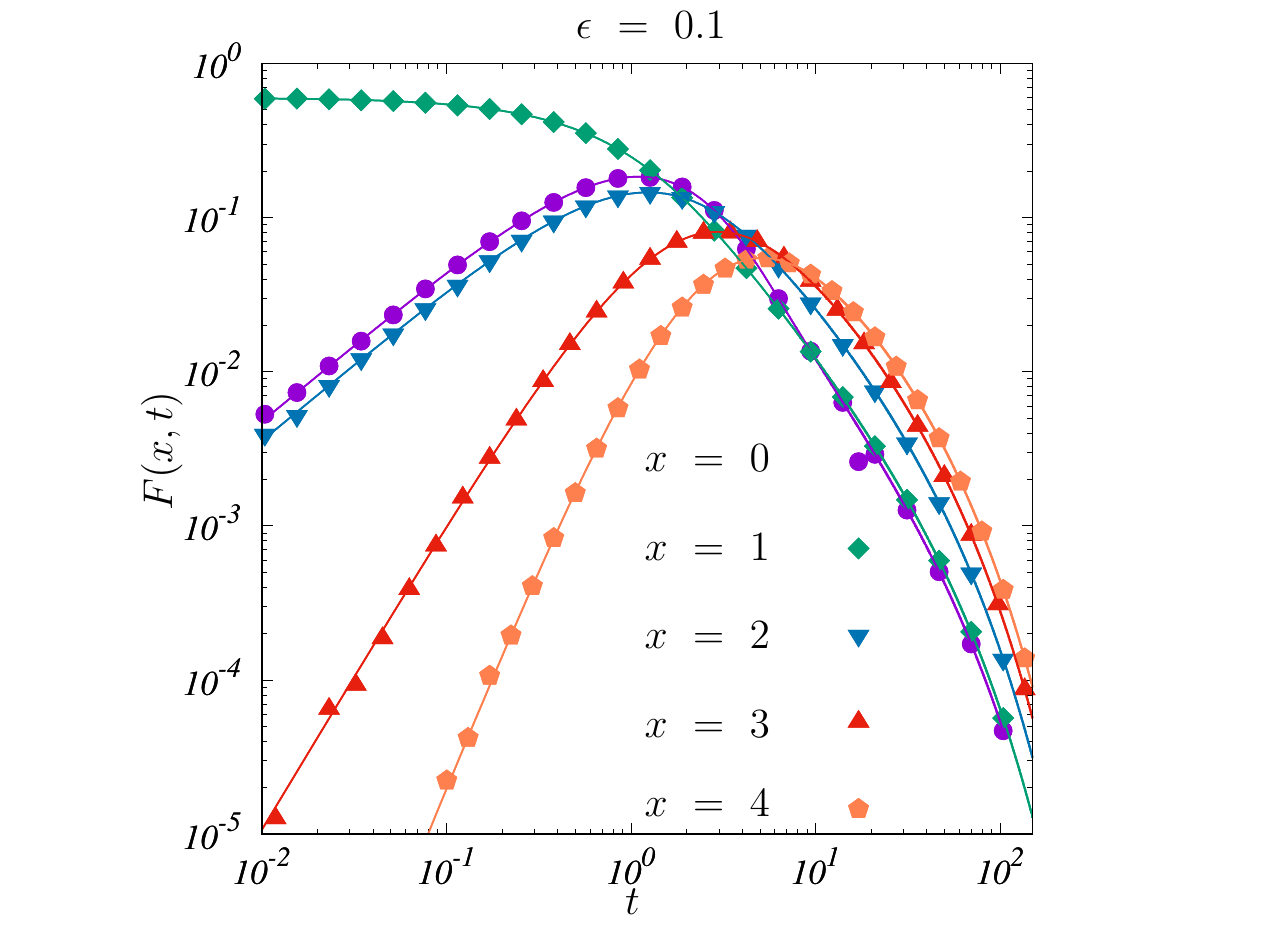}
        \caption{ }
    \end{subfigure}\hspace{-1.9 cm}
    \begin{subfigure}[t]{0.44\textwidth}
        \centering
        \includegraphics[width=\textwidth]{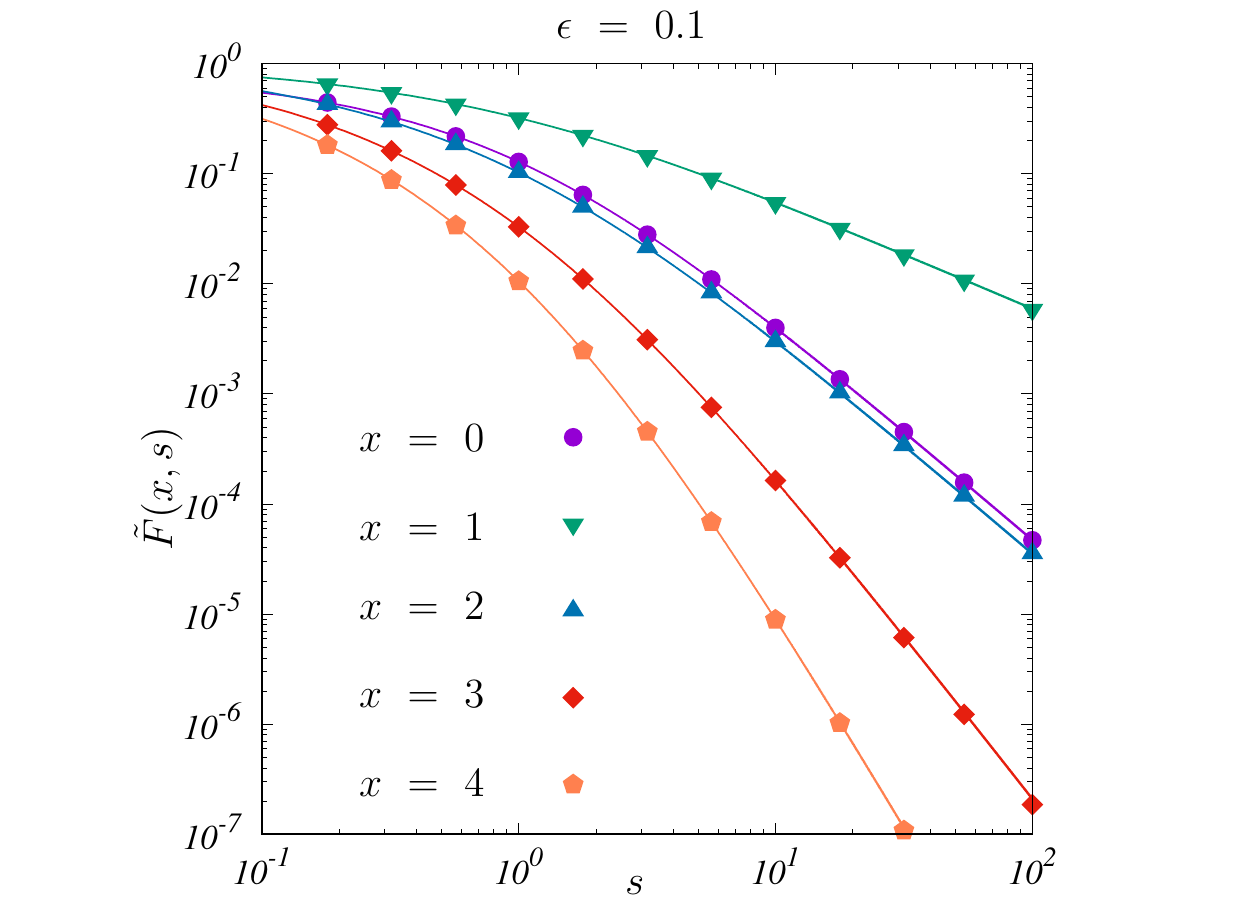}
                \caption{ }
    \end{subfigure}\hspace{-1.8 cm}
    \begin{subfigure}[t]{0.44\textwidth}
        \centering
        \includegraphics[width=\textwidth]{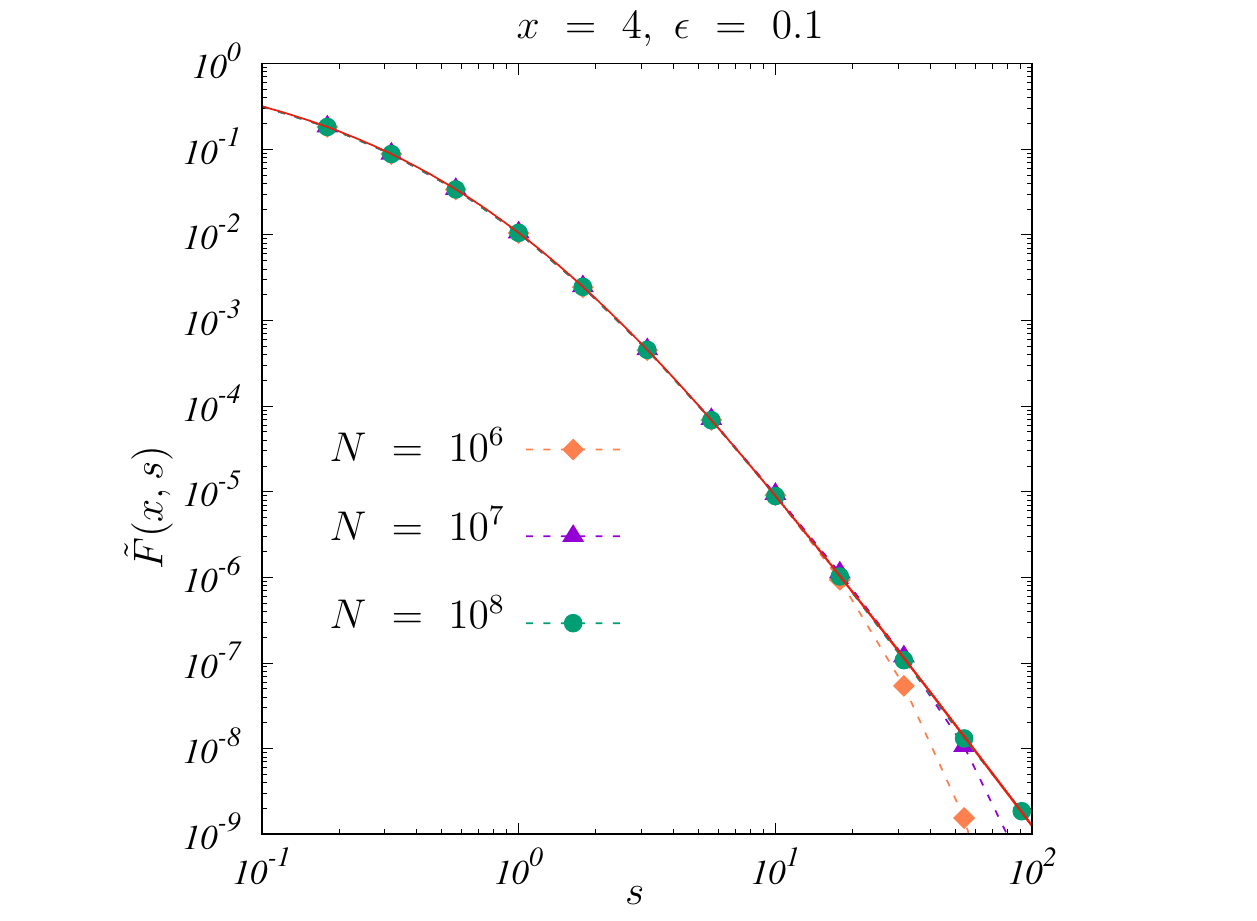}
        \caption{ }
    \end{subfigure}
    \hspace{-2 cm}
\caption{(a)~First passage probability density of a biased random walker $F(x,t)$, plotted as a function of time for different $x$. The solid lines correspond to the theoretical results in equations~(\ref{eq:a5}) and~(\ref{eq:a13}) whereas the points are from simulations. The fixed bias used is $\epsilon=0.1$. The simulation data is averaged over $10^8$ realisations.~(b) The Laplace transform of the first passage probability of a biased random walker $\tilde F(x,s)$, plotted as a function of $s$ for different $x$. The solid lines correspond to the theoretical results in equations~(\ref{eq:a4}) and~(\ref{eq:a8}) and the points are from simulations.~(c) Numerical simulation data (points) for $\tilde F(x,s)$ converge with the theoretical result (red solid line) more accurately as the number of realisations~$N$, is increased. The error in $F(x,t)$ due to finite sampling at short times appears as a correction to $\tilde F(x,s)$ in the large $s$ limit.  Here, $x$ is fixed to be $4$. } \label{fig:a1fxt}
\end{figure}
This implies that the biased random walk is transient in one dimension. If $\epsilon=0$ (symmetric random walk), then the walk is recurrent and $R=1$.
The Laplace transform in (\ref{eq:a8}) can be exactly inverted and it gives
\begin{small}
\begin{equation}
\label{eq:a13}
F(0,t)=2 \alpha  e^{-t} t \left(2+\pi  L_1 \left (2\sqrt{\alpha}t \right )\right) I_0\left(2  \sqrt{\alpha }t\right)-2 \sqrt{\alpha } e^{-t} \left (1+\sqrt{\alpha}\pi t L_0 \left(2\sqrt{\alpha}t \right)\right ) I_1\left(2  \sqrt{\alpha }t\right),
\end{equation}
\end{small}

where $\alpha=\left(\frac{1}{4}-\epsilon^2 \right)$.~Here,~$I_n(z)$ is the modified Bessel function of order $n$ defined previously~and~$L_n(z)$ is the modified Struve function~\cite{abramowitz1972handbook} of order $n$ which is a solution to the non-homogeneous Bessel differential equation $z^2\frac{d^2y}{dz^2}+z\frac{dy}{dz}-(z^2+n^2)y=\frac{4{\left( \frac{z}{2} \right)}^{n+1}}{\sqrt{\pi}\Gamma(n+\frac{1}{2})}$.
In series representation, these functions are given as
\begin{equation}
I_n(z)=\left(\frac{z}{2}\right)^{n} \sum _{k=0}^{\infty } \frac{\left(\frac{z}{2}\right)^{2 k}}{\Gamma (k+n+1) k!},
\end{equation}
and
\begin{equation}
    L_n(z)=\left(\frac{z}{2}\right)^{n+1} \sum _{k=0}^{\infty } \frac{\left(\frac{z}{2}\right)^{2
   k}}{\Gamma \left(k+\frac{3}{2}\right) \Gamma \left(k+n+\frac{3}{2}\right)}.
\end{equation}
The exact expression for the first return probability density for a one dimensional biased random walker given in~(\ref{eq:a13}) is one of the central results of this paper. The limits of this distribution are obtained as
\begin{equation} 
\label{symas2}
\lim_{t \rightarrow 0} F(0,t)= 2 \alpha t-2 \alpha t^2+\frac{1}{3} \left(3 \alpha +\alpha ^2\right)t^3+...
\end{equation}
and
\begin{small}
\begin{eqnarray} 
\label{symas2a}
\lim_{t \rightarrow \infty} F(0,t)&=& \frac{e^{(-1+2\sqrt{\alpha} ) t}}{2\sqrt{\pi } \alpha^{\frac{1}{4} }} \frac{1}{t^{\frac{3}{2}}}+\frac{9e^{(-1+2\sqrt{\alpha} ) t}}{32\sqrt{\pi } \alpha^{\frac{3}{4} }} \frac{1}{t^{\frac{5}{2}}}+\frac{345e^{(-1+2\sqrt{\alpha} ) t}}{1024\sqrt{\pi } \alpha^{\frac{5}{4} }} \frac{1}{t^{\frac{7}{2}}}+...
\end{eqnarray}
\end{small}

For a symmetric random walk $\left(\alpha=\frac{1}{4} \right)$, $F(0,t)$ reduces to
\begin{small}
\begin{equation} 
{F(0,t)}_{srw}=\frac{1}{2} e^{-t} \left( \left(2+\pi L_1(t) \right )t I_0(t)-\left (2+\pi  t
   L_0(t) \right ) I_1(t) \right ),
\end{equation}
\end{small}
and the limits are
\begin{equation} 
\label{symas}
\lim_{t \rightarrow 0} {F(0,t)}_{srw}=\frac{t}{2}-\frac{t^2}{2}+\frac{13 t^3}{48}+...
\end{equation}
\begin{small}
\begin{equation} 
\label{symasS}
\lim_{t \rightarrow \infty} {F(0,t)}_{srw}= \frac{1}{\sqrt{2 \pi}}\frac{1}{t^{\frac{3}{2}}}+\frac{9}{8 \sqrt{2 \pi }} \frac{1}{t^{5/2}}+\frac{345}{128 \sqrt{2 \pi }} \frac{1}{t^{7/2}}+...
\end{equation}
\end{small}

In figure~\ref{fig:a1fxt}, the analytical expressions in~(\ref{eq:a4}),~(\ref{eq:a5}),~(\ref{eq:a8}) and~(\ref{eq:a13}) for the first passage probability density of a biased random walker in one dimension are compared with the numerical simulations of the same. In order to extract $\tilde F (x,s)$, we perform numerical Laplace transformation of the data for $F(x,t)$. Since it is difficult to resolve the small time first passage probabilities exactly to arbitrary accuracy by simulations, the numerical Laplace transform exhibits a finite error in the large $s$ limit. This error reduces by increasing the number of realisations being averaged.

\section{Symmetric random walk in two dimensions}
\label{sec:3}
We next consider the motion of a symmetric random walker on a two dimensional square lattice where the lattice points are labelled by variables $(x,y)$. The simplest case where only nearest neighbour hops are allowed is studied. Let $P(x,y,t)$ be the probability for the random walker to be at lattice position $(x,y)$ at time $t$ given that $P(x,y,t=0)=\delta_{x,0}\delta_{y,0}$. The translational rate for a symmetric random walker along all the four lattice directions is $\frac{1}{4}$. One can write the equation for the evolution of $P(x,y,t)$ as
\begin{small}
\begin{equation}
\label{eq:a12d}
 \frac{\partial P(x,y,t)}{\partial t}=\frac{1}{4} \left(P(x-1,y,t)+ P(x+1,y,t)+P(x,y-1,t)+P(x,y+1,t)\right)-P(x,y,t).
\end{equation}
\end{small}
\subsection{Occupation probability}
\label{subsection:3:1}
Let us define the Fourier-Laplace transform of the occupation probability as
\begin{equation}
\label{ft2dsq}
\tilde P(k_x,k_y,s)=\int_0^{\infty}\sum_{x=-\infty}^{\infty}\sum_{y=-\infty}^{\infty}e^{i(k_xx+k_yy)-st}P(x,y,t)dt.
\end{equation}

From (\ref{eq:a12d}), we get
\begin{equation} 
\tilde P(k_x,k_y,s)=\frac{2}{2 (1+s)-\cos k_x-\cos k_y}.
\end{equation}
Doing a Fourier inversion of the above equation gives the Laplace transform of the occupation probability. That is,
\begin{equation}
\label{latticegreenfun}
\tilde P(x,y,s)=\frac{1}{4 {\pi}^2}\int_{-\pi}^{\pi}\int_{-\pi}^{\pi}e^{-i(k_xx+k_yy)}P(k_x,k_y,s)dk_x dk_y.
\end{equation}

Integrals of this kind appear when we deal with the Green's functions for various lattice structures in different dimensions. The Green's function is the kernel of the discrete Laplacian operator~($\nabla^2$) on the $d$ dimensional lattice. Although these lattice Green's functions~\cite{katsura1971latticea,katsura1971latticem,morita1971calculation,katsura1971lattice,horiguchi1972lattice,ray2014green,guttmann2010lattice} have been computed in closed form in terms of generalised hypergeometric functions~\cite{bateman1953higher} and other special functions, the fact that these are exactly the same integrals that appear in the random walk context is still not widely identified. For example, the integral in~(\ref{latticegreenfun}) is exactly the lattice Green's function for a square lattice~\cite{katsura1971latticea}. For references on the applications of the lattice Green's functions in random walk studies, see~\cite{hughes1995random,moran1973gaussian,maassarani2000series}. Other examples for the applications of lattice Green's functions in varying physical situations can be found in~\cite{venezian1994resistance,cserti2000application,atkinson1999infinite,ivashkevich1998introduction}. This paper utilises the already known results for lattice Green's functions in the literature to compute the double integral in~(\ref{latticegreenfun}).  Katsura and Inawashiro have exactly computed the square lattice Green's function in~\cite{katsura1971latticea}. Using this, we obtain
\begin{small}
\begin{equation}
\label{eq:x}
\tilde P(x,y,s)=\frac{ \,
   _4F_3\left(\frac{1+|x|+|y|}{2},\frac{1+|x|+|y|}{2},\frac{2+|x|+|y|}{2},\frac{2+|x|+|y|}{2};1+|x|,1+|y|,1+|x|+|y|;\frac{1}{(1+s)^2}\right)}{2^{2 (|x|+|y|)} (1+s)^{1+|x|+|y|} \frac{|x|! |y|!}{(|x|+|y|)!}}.
\end{equation}
\end{small}
\begin{figure}[t!]
    \centering
    \hspace{-1.9 cm}
        \begin{subfigure}[t]{0.44\textwidth}
        \centering
        \includegraphics[width=\textwidth]{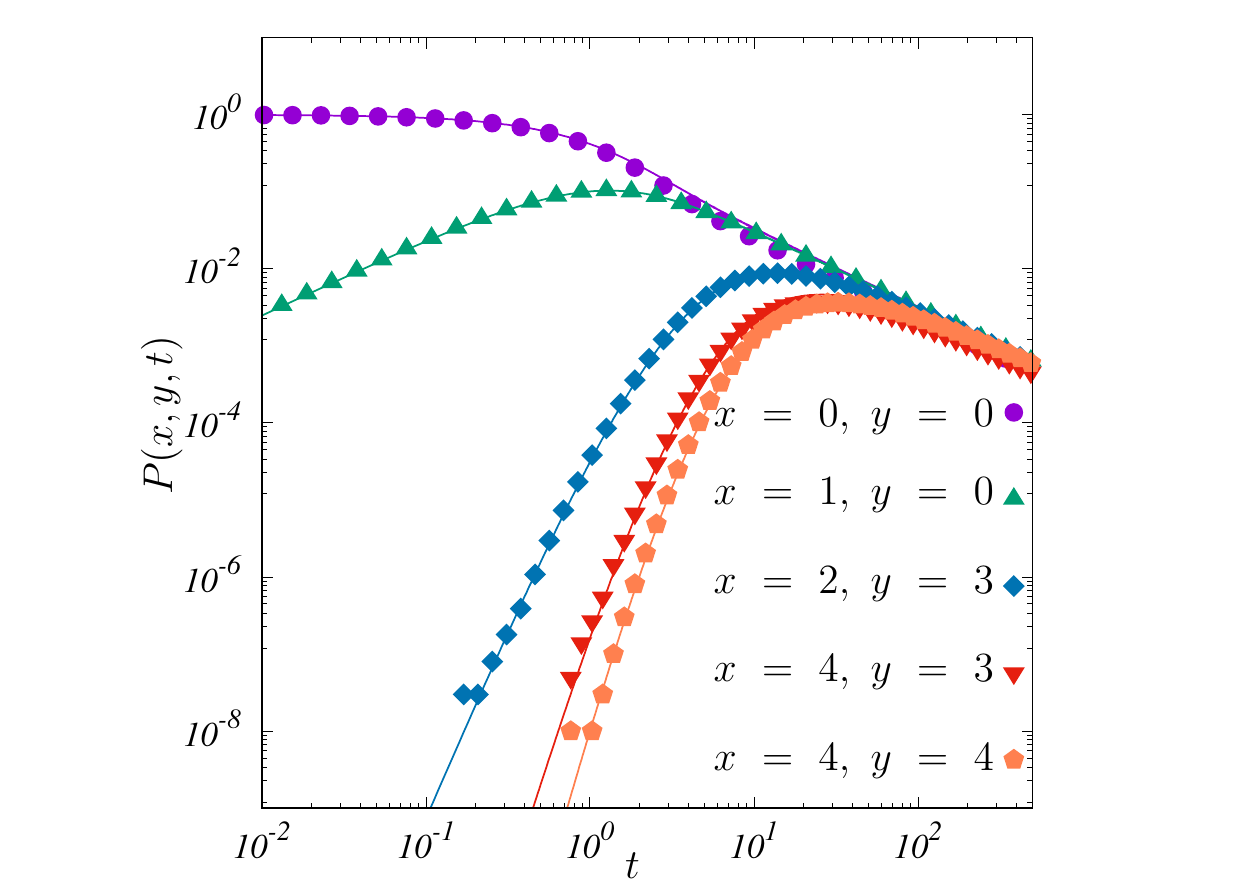}
        \caption{ }\label{fig:a1pxt2a} 
    \end{subfigure}\hspace{-1.9 cm}
    \begin{subfigure}[t]{0.44\textwidth}
        \centering
        \includegraphics[width=\textwidth]{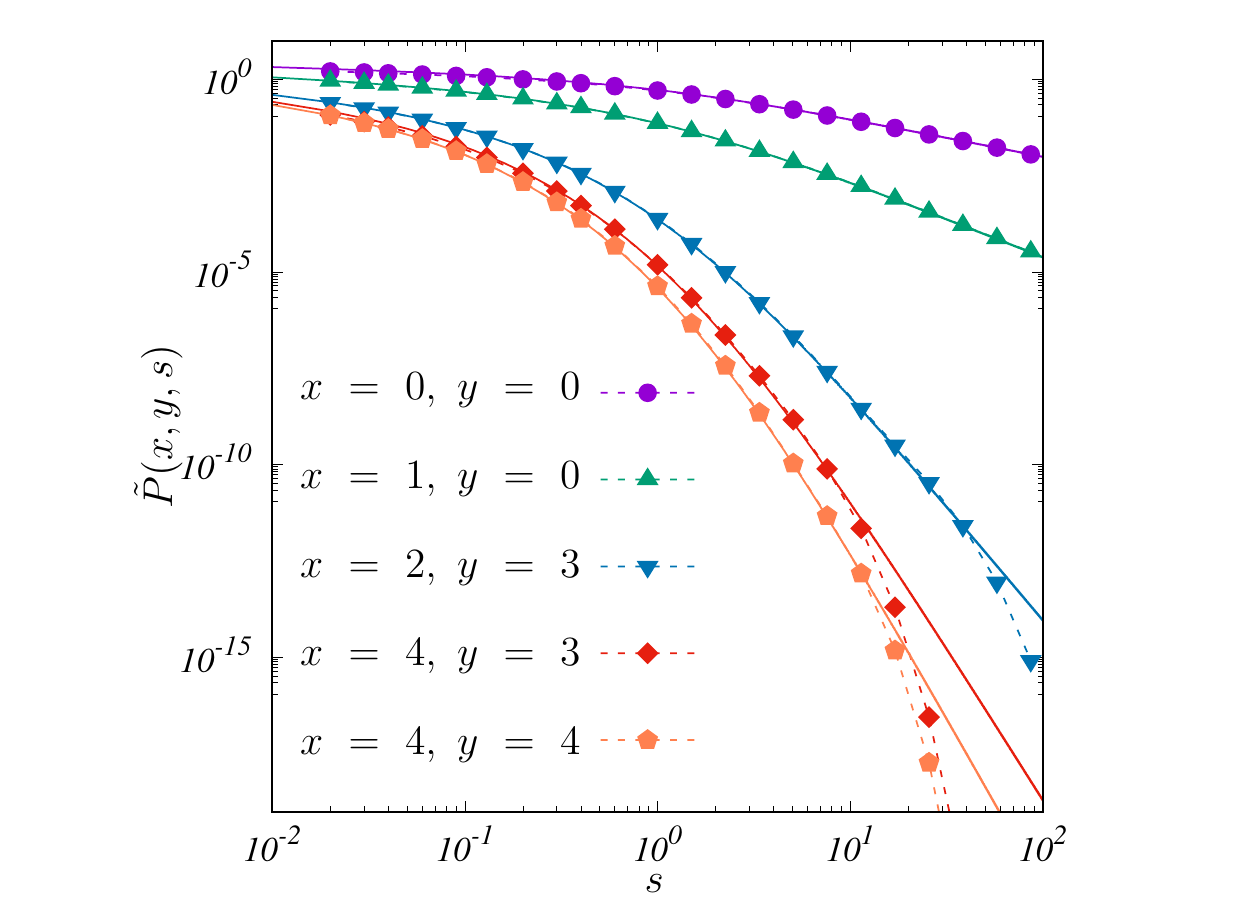}
        \caption{ }\label{fig:a1pxt2b} 
    \end{subfigure}\hspace{-1.9 cm}
    \begin{subfigure}[t]{0.44\textwidth}
        \centering
        \includegraphics[width=\textwidth]{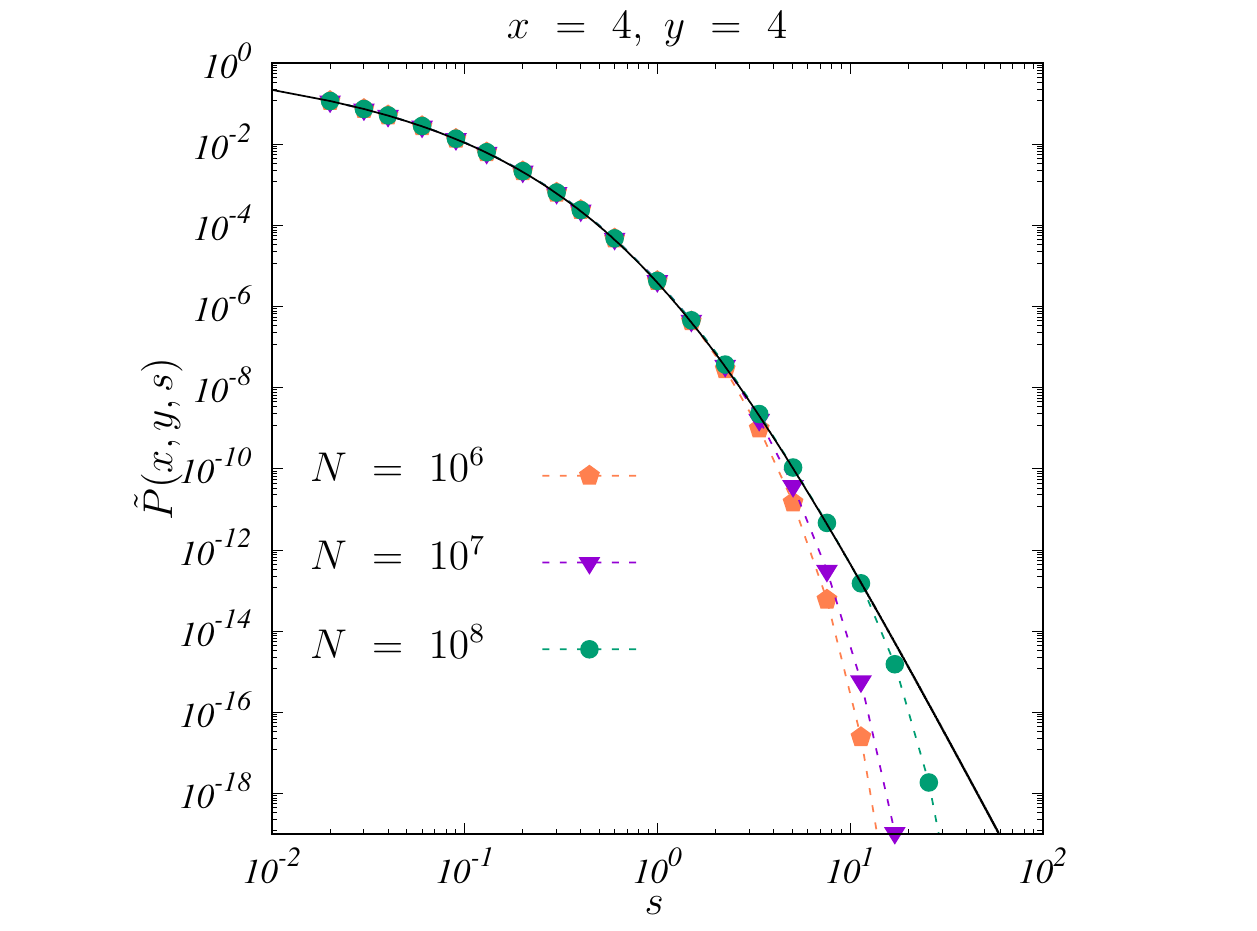}
        \caption{ }    \label{fig:a1pxt2c}
    \end{subfigure}
    \hspace{-2 cm}
\caption{(a) Occupation probability of a symmetric random walker $P(x,y,t)$, in two dimensions plotted as a function of time for different $x$, $y$. The solid lines correspond to the theoretical result in (\ref{eq:w}) whereas the points are from simulations. The simulation data is averaged over $10^8$ realisations.~(b) The Laplace transform of the occupation probability of a symmetric random walker in two dimensions $\tilde P(x,y,s)$, plotted as a function of $s$ for different $x$, $y$. The solid lines correspond to the theoretical result in (\ref{eq:x}) and the points are from simulations.~(c) Numerical simulation data (points) for $\tilde P(x,y,s)$ converge with the theoretical result (black solid line) more accurately as the number of realisations is increased. The error in $P(x,y,t)$ due to finite sampling at short times appears as a correction to $\tilde P(x,y,s)$ in the large $s$ limit. Here, $x$ and $y$ are fixed to be $4$.} \label{fig:a1pxt2}
\end{figure}
The general hypergeometric function $\, _pF_q \left (a_1,a_2,...,a_p;b_1,b_2,...,b_q;z \right )$~\cite{bell2004special} is defined as
\begin{equation}
\, _pF_q \left (a_1,a_2,...,a_p;b_1,b_2,...,b_q;z \right )=\sum _{k=0}^{\infty } \frac{{(a_1)}_k \ldots  {(a_p)}_k z^k}{{(b_1)}_k \ldots  {(b_q)}_k k!},
\end{equation}
where ${(a)}_k$ [and b$(k)$] is the rising factorial or the Pochhammer symbol given as
\begin{equation}
   {(a)}_k=
 \frac{\Gamma(a+k)}{\Gamma(a)},~\forall~k \in \mathbb{Z^*}.
\end{equation}
For $x=y$, the integration yields,
\begin{equation}
\label{hh}
\tilde P(x,s)=\frac{\Gamma \left(\frac{1}{2}+|x|\right) \, _2F_1\left(\frac{1+2|x|}{2},\frac{1+2|x|}{2};1+2
   |x|;\frac{1}{(1+s)^2}\right)}{2^{2 |x|} (1+s)^{1+2 |x|} \sqrt{\pi } \Gamma (1+|x|)},
\end{equation}
where $\tilde P(x,s)=\tilde P(x,x,s)$ from which we get
\begin{equation} 
\label{eq:y}
\tilde P(0,s)=\frac{2 K\left(\frac{1}{(1+s)^2}\right)}{\pi  (1+s)}.
\end{equation}
$\Gamma$ is the Gamma function and $K$ is the elliptic integral of the first kind defined as
\begin{equation} 
K(m)=\int_{0}^{\frac{\pi}{2}} \frac{1}{\sqrt{1-m {\sin}^2(\theta )}} \, d\theta.
\end{equation}
The limits of the Laplace transform in (\ref{eq:y}) are found to be
\begin{small}
\begin{eqnarray} 
\label{srw2ds0}
\lim_{s \rightarrow 0} \tilde P (0,s)&=&-\frac{\log \left(\frac{s}{8}\right)}{\pi }+\frac{s \left(1+\log \left(\frac{s}{8}\right)\right)}{2 \pi }-\frac{s^2 \left(7+5 \log \left(\frac{s}{8}\right)\right)}{16 \pi }+...,
\end{eqnarray}
\end{small}
and
\begin{equation} 
\label{srw2ds1}
\lim_{s \rightarrow \infty} \tilde P (0,s)\approx \frac{1}{s}-\frac{1}{s^2}+\frac{5}{4 s^3}-\frac{7}{4 s^4}+....
\end{equation}
(\ref{eq:x}) on Laplace inversion gives the known result \cite{balakrishnan1983some},
\begin{equation} 
\label{eq:w}
 P(x,y,t)=e^{-t} {I_{|x|}}\left(\frac{t}{2}\right){I_{|y|}}\left(\frac{t}{2}\right).
\end{equation}

The comparison of the theoretical results for the occupation probability in two dimensions (\ref{eq:x}),~(\ref{eq:w}) for different combinations of $x$ and $y$ with numerical simulation data is presented in figure~\ref{fig:a1pxt2}. For the lattice points close to the origin, numerical data fits well with the theoretical prediction. However, the small time occupation probabilities of the lattice points farther away from the origin are not well resolved due to finite sampling. This appears as a correction to the Laplace transform in the large $s$ limit (see figure~\ref{fig:a1pxt2b}). This error can be reduced by increasing the number of realisations being averaged. The convergence of the simulation data for $\tilde P (x,y,s)$ to the theoretical prediction with an increase in the number of realisations is shown in figure~\ref{fig:a1pxt2c}.
\subsection{First passage probability}
\label{subsection:3:2}
Similar to the calculation in one dimension, one can write recursion relations connecting the Laplace transforms of the occupation probability and the first passage probability density in two dimensions. For $x=y=0$, we get
\begin{equation}
\label{eq:v}
\tilde F(0,s)=1-\frac{1}{\tilde P (0,s)(s+1)},
\end{equation}
where $\tilde F(0,s)=\tilde F(0,0,s)$ and $\tilde P(0,s)=\tilde P(0,0,s)$.
For non zero $x,~y$, we get
\begin{equation}
\label{rec2d}
\tilde F(x,y,s)=\frac{\tilde P(x,y,s)}{\tilde P (0,s)}, ~\forall~x,~y \neq 0.
\end{equation}
Using (\ref{eq:x}) and (\ref{eq:y}) in (\ref{rec2d}) yields
\begin{small}
\begin{equation}
\label{hf1}
\tilde F(x,y,s)=\frac{ \,
   _4F_3\left(\frac{1+|x|+|y|}{2},\frac{1+|x|+|y|}{2},\frac{2+|x|+|y|}{2},\frac{2+|x|+|y|}{2};1+|x|,1+|y|,1+|x|+|y|;\frac{1}{(1+s)^2}\right)}{2 K\left(\frac{1}{(1+s)^2}\right)2^{2 (|x|+|y|)} (1+s)^{|x|+|y|} \frac{|x|! |y|!}{\pi (|x|+|y|)!}}.
\end{equation}
\end{small}
(\ref{hf1}) is the exact analytic expression for the  characteristic function of the first passage time distribution of a symmetric random walker on a two dimensional square lattice. This expression holds for any site other than the origin.
The corresponding expression for diagonal sites is greatly simplified by setting $x=y$ and the characteristic function turns out to be
\begin{small}
\begin{equation}
\label{hf}
\tilde F(x,s)=\frac{\pi\Gamma \left(\frac{1}{2}+x\right) \, _2F_1\left(\frac{1}{2}+x,\frac{1}{2}+x;1+2
   x;\frac{1}{(1+s)^2}\right)}{2^{2 x} (1+s)^{2 x} \sqrt{\pi } \Gamma (1+x)2 K\left(\frac{1}{(1+s)^2}\right)},
\end{equation}
\end{small}
where $\tilde F(x,s)=\tilde F(x,x,s)$. 
The characteristic function for the time of first return to the origin ($x=y=0$) is found by using~(\ref{eq:y}) in~(\ref{eq:v}). This gives
\begin{equation}
\label{eq:z}
\tilde F(0,s)=1-\frac{\pi }{2 K\left(\frac{1}{(1+s)^2}\right)}.
\end{equation}
The limits to (\ref{eq:z}) are found to be
\begin{equation} 
\label{srw2ds2}
\lim_{s \rightarrow 0} \tilde F (0,s)= 1+ \frac{\pi }{\log \left(\frac{s}{8}\right)}+\frac{\pi  s (1-\log \left(\frac{s}{8} \right))}{2 \log ^2\left(\frac{s}{8}\right)}+...,
\end{equation}
\begin{equation} 
\label{srw2ds3}
\lim_{s \rightarrow \infty} \tilde F (0,s)= \frac{1}{4 s^2}-\frac{1}{2 s^3}+\frac{53}{64 s^4}-\frac{21}{16 s^5}+....
\end{equation}
The asymptotic behaviour of $F(0,t)$ can be deduced from (\ref{srw2ds2}) by doing a term by term Laplace inversion. For the leading order term, we have
\begin{eqnarray}
\label{reca3}
\lim_{t \rightarrow \infty}F(0,t)&=&\frac{\pi}{t {{\log}^2 \left(t\right)}},
\end{eqnarray}
which matches with the asymptotic time dependence in~\cite{redner2001guide}.

\begin{figure}[t!]
    \centering
        \begin{subfigure}[t]{0.44\textwidth}
        \centering
        \includegraphics[width=\textwidth]{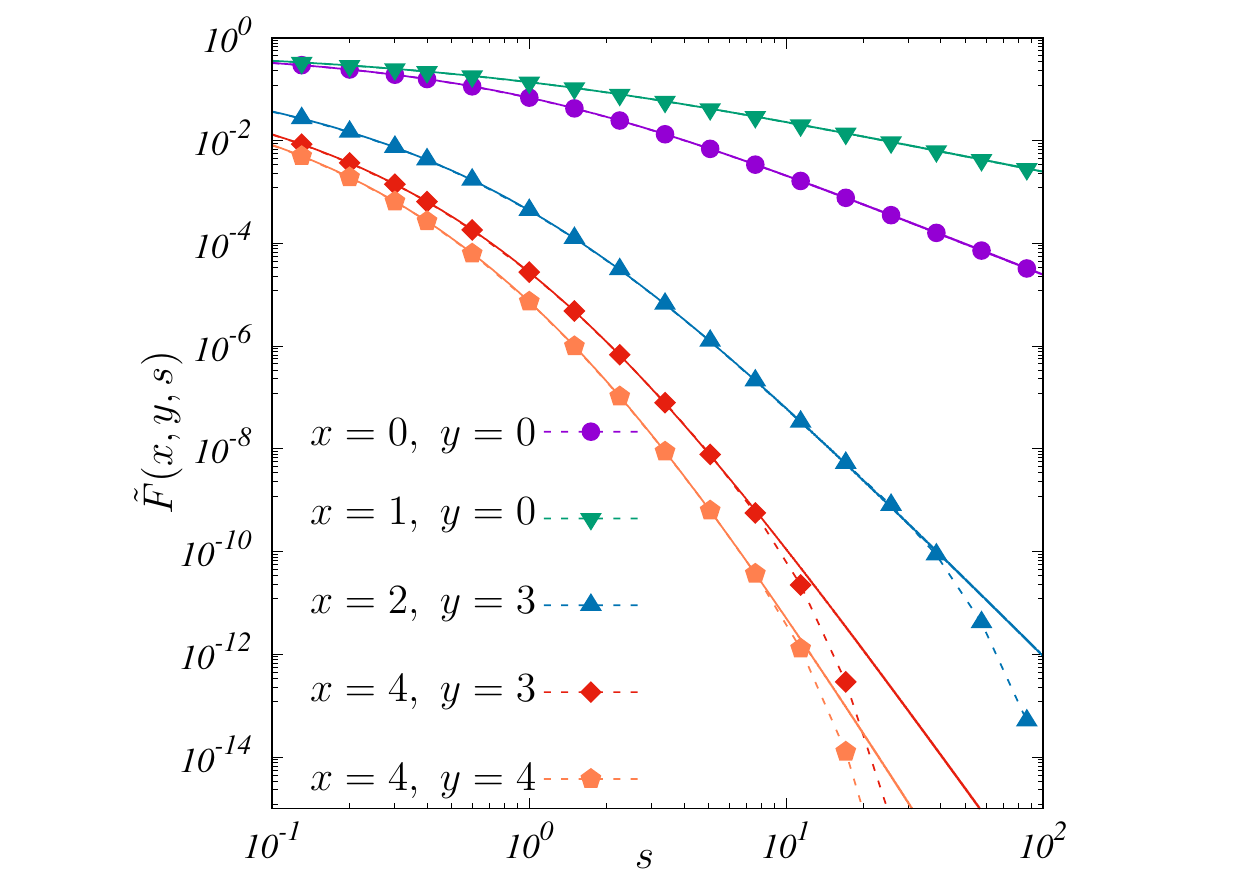}
        \caption{ }\label{fig:a1fxt2a}
    \end{subfigure}\hspace{-1.5 cm}
    ~ 
    \begin{subfigure}[t]{0.44\textwidth}
        \centering
        \includegraphics[width=\textwidth]{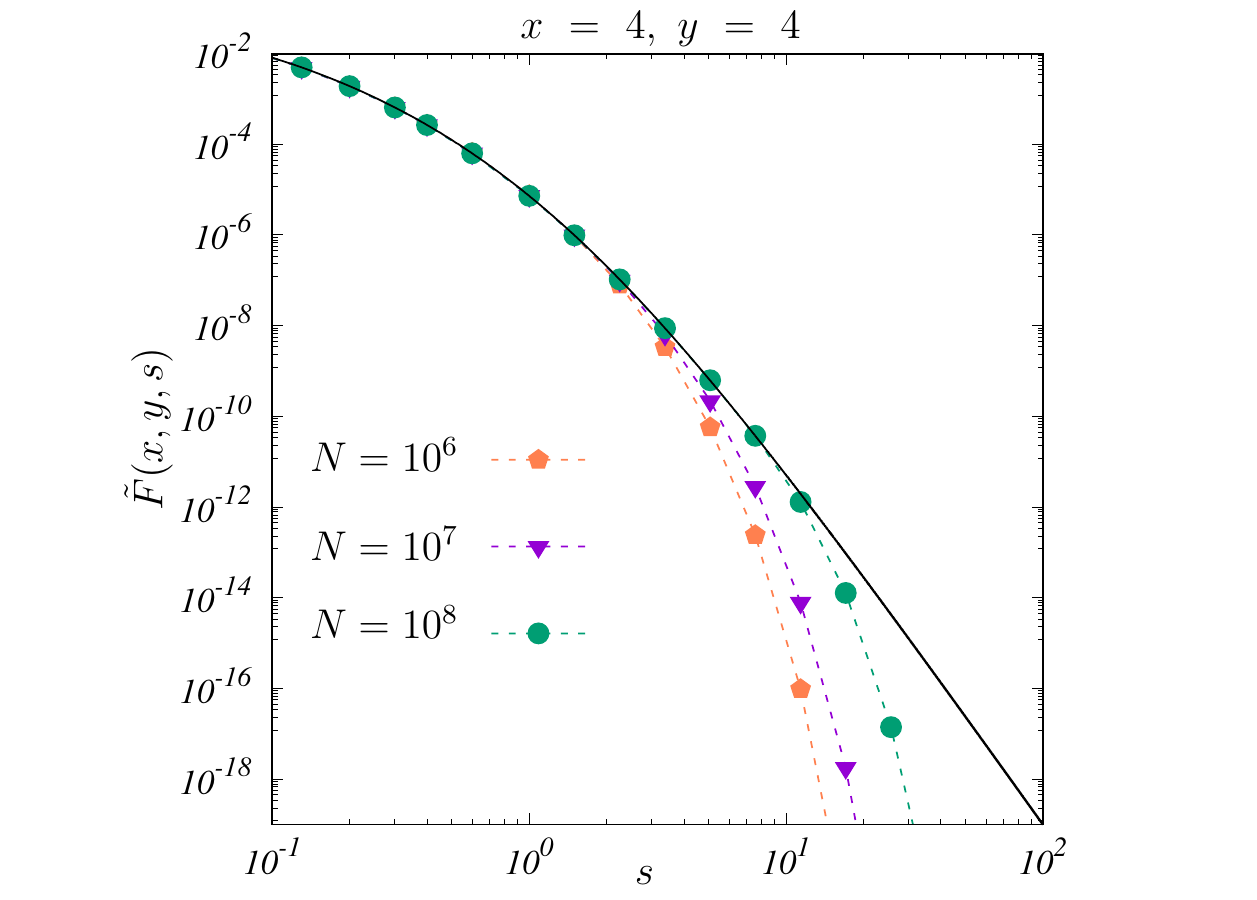}
        \caption{ }\label{fig:a1fxt2b}
    \end{subfigure}
\caption{ 
(a)~The characteristic function of the first passage time distribution of a symmetric random walker on a two dimensional square lattice $\tilde F(x,y,s)$, plotted as a function of $s$ for different $x$, $y$. The solid lines correspond to the theoretical result in (\ref{hf1}) whereas the points are from simulations. The simulation data is averaged over $10^8$ realisations.~(b) Numerical simulation data (points) for $\tilde F(x,y,s)$ converge with the theoretical result (black solid line) more accurately as the number of realisations is increased. The error in $F(x,y,t)$ due to finite sampling appears as a correction to $\tilde F(x,y,s)$ in the large $s$ limit. Here, $x$ and $y$ are fixed to be $4$.} \label{fig:a1fxt2}
\end{figure}
The characteristic function $\tilde F(x,y,s)$, in~(\ref{hf1}) is compared with numerical simulation data in figure~\ref{fig:a1fxt2}. Numerically, this is done by taking a numerical Laplace transform of the data for $F(x,y,t)$. For the sites closer to the origin, finite sampling gives a very good fit with the theoretical prediction. The error in $\tilde F(x,y,s)$ due to finite sampling in the large $s$ limit increases as we go farther from the origin. However, this can be reduced by increasing the number of realisations being averaged~(see figure~\ref{fig:a1fxt2b}).

\section{Summary and conclusions}
\label{ccl}
In this paper, we have investigated the first passage properties of continuous time Markov processes on one and two dimensional infinite lattices. We have derived exact closed forms for the first passage probability density in one and two dimensions. First, we derived an exact expression for the probability of first return to the origin of a biased random walker in one dimension in terms of modified Bessel and Struve functions. Next, we obtained an exact expression for the characteristic function or the Laplace transform of the probability of first passage to an arbitrary site of a symmetric random walker on a two dimensional square lattice. This primary result for the characteristic function follows from the mapping to the square lattice Green's function. As the mapping to lattice Green's functions is quite general, it would be interesting to extend this mapping to other lattice structures as well as to higher dimensions~\cite{katsura1971latticea,katsura1971latticem,morita1971calculation,katsura1971lattice,horiguchi1972lattice,ray2014green,guttmann2010lattice}.
\section{Acknowledgments}

Useful discussions with Kabir Ramola, Mustansir Barma, Dipanjan Mandal, Roshan Maharana, Vishnu V.~Krishnan, Pappu Acharya, Debankur Das and Soham Mukhopadhyay are gratefully acknowledged. This project was funded by intramural funds at TIFR Hyderabad from the Department of Atomic Energy (DAE).
\section*{References}
\bibliography{bibtex.bib}{}

\providecommand{\newblock}{}
\begin{thebibliography}{10}
\expandafter\ifx\csname url\endcsname\relax
  \def\url#1{{\tt #1}}\fi
\expandafter\ifx\csname urlprefix\endcsname\relax\def\urlprefix{URL }\fi
\providecommand{\eprint}[2][]{\url{#2}}

\bibitem{chou2014first}
Chou T and D'Orsogna M~R 2014 First passage problems in biology {\em
  First-passage phenomena and their applications\/} (World Scientific) pp
  306--345

\bibitem{kenwright2012first}
Kenwright D~A, Harrison A~W, Waigh T~A, Woodman P~G and Allan V~J 2012 {\em
  Physical Review E\/} {\bf 86} 031910

\bibitem{condamin2008probing}
Condamin S, Tejedor V, Voituriez R, B{\'e}nichou O and Klafter J 2008 {\em
  Proceedings of the National Academy of Sciences\/} {\bf 105} 5675--5680

\bibitem{szabo1980first}
Szabo A, Schulten K and Schulten Z 1980 {\em The Journal of chemical physics\/}
  {\bf 72} 4350--4357

\bibitem{park2003reaction}
Park S, Sener M~K, Lu D and Schulten K 2003 {\em The Journal of chemical
  physics\/} {\bf 119} 1313--1319

\bibitem{liu2017anchoring}
Liu H, Liao C~Y, Ko J~Y and Lih J~S 2017 {\em Physica A: Statistical Mechanics
  and its Applications\/} {\bf 477} 114--127

\bibitem{zhang2009first}
Zhang D and Melnik R~V 2009 {\em Applied Stochastic Models in Business and
  Industry\/} {\bf 25} 565--582

\bibitem{chicheportiche2014some}
Chicheportiche R and Bouchaud J~P 2014 Some applications of first-passage ideas
  to finance {\em First-Passage Phenomena and Their Applications\/} (World
  Scientific) pp 447--476

\bibitem{polya1921aufgabe}
P{\'o}lya G 1921 {\em Mathematische Annalen\/} {\bf 84} 149--160

\bibitem{montroll1965random}
Montroll E~W and Weiss G~H 1965 {\em Journal of Mathematical Physics\/} {\bf 6}
  167--181

\bibitem{montroll1979enriched}
Montroll E~W and West B~J 1979 {\em Fluctuation phenomena\/} {\bf 66} 61

\bibitem{kutner2017continuous}
Kutner R and Masoliver J 2017 {\em The European Physical Journal B\/} {\bf 90}
  1--13

\bibitem{mainardi2020advent}
Mainardi F 2020 {\em Mathematics\/} {\bf 8} 641

\bibitem{siegert1951first}
Siegert A~J 1951 {\em Physical Review\/} {\bf 81} 617

\bibitem{burkhardt2014first}
Burkhardt T~W 2014 First passage of a randomly accelerated particle {\em
  First-Passage Phenomena and Their Applications\/} (World Scientific) pp
  21--44

\bibitem{benichou2014first}
B{\'e}nichou O and Voituriez R 2014 First-passage times of intermittent random
  walks {\em First-passage Phenomena And Their Applications\/} (World
  Scientific) pp 70--95

\bibitem{khantha1983first}
Khantha M and Balakrishnan V 1983 {\em Pramana\/} {\bf 21} 111--122

\bibitem{balakrishnan1983first}
Balakrishnan V and Khantha M 1983 {\em Pramana\/} {\bf 21} 187--200

\bibitem{redner2001guide}
Redner S 2001 {\em A guide to first-passage processes\/} (Cambridge university
  press)

\bibitem{feller2008introduction}
Feller W 2008 {\em An introduction to probability theory and its applications,
  vol 2\/} (John Wiley \& Sons)

\bibitem{balakrishnan1983some}
Balakrishnan V 1983 {\em Rendiconti del Seminario Matematico e Fisico di
  Milano\/} {\bf 53} 273--284

\bibitem{balakrishnan1988first}
Balakrishnan V, Van~den Broeck C and H{\"a}nggi P 1988 {\em Physical Review
  A\/} {\bf 38} 4213

\bibitem{balakrishnan1981two}
Balakrishnan V and Venkataraman G 1981 {\em Pramana\/} {\bf 16} 109--130

\bibitem{prasad1984biased}
Prasad M and Unnikrishnan K 1984 {\em Physica A: Statistical Mechanics and its
  Applications\/} {\bf 127} 659--666

\bibitem{prados1997dynamical}
Prados A, Brey J and S{\'a}nchez-Rey B 1997 {\em Journal of statistical
  physics\/} {\bf 89} 709--734

\bibitem{bortz1975new}
Bortz A~B, Kalos M~H and Lebowitz J~L 1975 {\em Journal of Computational
  Physics\/} {\bf 17} 10--18

\bibitem{voter2007introduction}
Voter A~F 2007 Introduction to the kinetic monte carlo method {\em Radiation
  effects in solids\/} (Springer) pp 1--23

\bibitem{haus1987diffusion}
Haus J~W and Kehr K~W 1987 {\em Physics Reports\/} {\bf 150} 263--406

\bibitem{abramowitz1972handbook}
Abramowitz M and Stegun I~A 1972 {\em Handbook of mathematical functions with
  formulas, graphs, and mathematical tables\/} vol~55 (US Government printing
  office)

\bibitem{katsura1971latticea}
Katsura S and Inawashiro S 1971 {\em Journal of Mathematical Physics\/} {\bf
  12} 1622--1630

\bibitem{katsura1971latticem}
Katsura S, Inawashiro S and Abe Y 1971 {\em Journal of Mathematical Physics\/}
  {\bf 12} 895--899

\bibitem{morita1971calculation}
Morita T and Horiguchi T 1971 {\em Journal of Mathematical Physics\/} {\bf 12}
  986--992

\bibitem{katsura1971lattice}
Katsura S, Morita T, Inawashiro S, Horiguchi T and Abe Y 1971 {\em Journal of
  Mathematical Physics\/} {\bf 12} 892--895

\bibitem{horiguchi1972lattice}
Horiguchi T 1972 {\em Journal of Mathematical Physics\/} {\bf 13} 1411--1419

\bibitem{ray2014green}
Ray K 2014 {\em arXiv preprint arXiv:1409.7806\/}

\bibitem{guttmann2010lattice}
Guttmann A~J 2010 {\em Journal of Physics A: Mathematical and Theoretical\/}
  {\bf 43} 305205

\bibitem{bateman1953higher}
Bateman H 1953 {\em Higher transcendental functions [volumes i-iii]\/} vol~1
  (McGraw-Hill Book Company)

\bibitem{hughes1995random}
Hughes B 1995 {\em Random Walks and Random Environments, vol 1\/} (Oxford
  University Press)

\bibitem{moran1973gaussian}
Moran P~A 1973 {\em Journal of Applied Probability\/} {\bf 10} 54--62

\bibitem{maassarani2000series}
Maassarani Z 2000 {\em Journal of Physics A: Mathematical and General\/} {\bf
  33} 5675

\bibitem{venezian1994resistance}
Venezian G 1994 {\em American Journal of Physics\/} {\bf 62} 1000--1004

\bibitem{cserti2000application}
Cserti J 2000 {\em American Journal of Physics\/} {\bf 68} 896--906

\bibitem{atkinson1999infinite}
Atkinson D and Van~Steenwijk F 1999 {\em American Journal of Physics\/} {\bf
  67} 486--492

\bibitem{ivashkevich1998introduction}
Ivashkevich E and Priezzhev V~B 1998 {\em Physica A: Statistical Mechanics and
  its Applications\/} {\bf 254} 97--116

\bibitem{bell2004special}
Bell W~W 2004 {\em Special functions for scientists and engineers\/} (Courier
  Corporation)

\end{thebibliography}
\bibliographystyle{iopart-num}
\end{document}